\documentclass[11pt]{article}
\usepackage{times}
\usepackage{geometry}
\usepackage[utf8]{inputenc}
\usepackage{enumitem,amssymb}
\usepackage{ragged2e}
\usepackage{graphicx}
\usepackage{comment}
\usepackage{multicol}
\usepackage[usenames]{xcolor} %used for font color   
\usepackage{wrapfig}

\definecolor{xlinkcolor}{cmyk}{1,1,0,0}
\usepackage{url}
\usepackage[
 colorlinks=true,    % false: boxed links; true: colored links 
 linkcolor=xlinkcolor,     % color of internal links            
 citecolor=xlinkcolor,     % color of links to bibliography
 filecolor=xlinkcolor,  % color of file links 
 urlcolor=xlinkcolor,      % color of external link
 final=true
]{hyperref}
\usepackage[super,sort&compress]{natbib}
\usepackage{enumitem}
\setenumerate{itemsep=0mm}

\begin{document}
\begin{raggedright} 
% part of template, but does not look good
\huge
Snowmass2021 - Letter of Interest \hfill \\[+1em]
\textit{Cosmology Intertwined IV: The Age of the Universe and its Curvature} \hfill \\[+1em]
\end{raggedright}

\normalsize

\noindent {\large \bf Thematic Areas:}  (check all that apply $\square$/$\blacksquare$)

\noindent $\blacksquare$ (CF1) Dark Matter: Particle Like \\
\noindent $\square$ (CF2) Dark Matter: Wavelike  \\ 
\noindent $\square$ (CF3) Dark Matter: Cosmic Probes  \\
\noindent $\blacksquare$ (CF4) Dark Energy and Cosmic Acceleration: The Modern Universe \\
\noindent $\square$ (CF5) Dark Energy and Cosmic Acceleration: Cosmic Dawn and Before \\
\noindent $\square$ (CF6) Dark Energy and Cosmic Acceleration: Complementarity of Probes and New Facilities \\
\noindent $\blacksquare$ (CF7) Cosmic Probes of Fundamental Physics \\
\noindent $\square$ (Other) {\it [Please specify frontier/topical group]} \\

\noindent {\large \bf Contact Information:}\\
Eleonora Di Valentino (JBCA, University of Manchester, UK) [eleonora.divalentino@manchester.ac.uk]\\
%Collaboration (optional): \\

\noindent {\large \bf Authors:}  \\[+1em]
Eleonora Di Valentino (JBCA, University of Manchester, UK)\\
Luis A. Anchordoqui (City University of New York, USA)\\
\"{O}zg\"{u}r Akarsu (Istanbul Technical University, Istanbul, Turkey) \\
Yacine Ali-Haimoud (New York University, USA)\\
Luca Amendola (University of Heidelberg, Germany)\\
Nikki Arendse (DARK, Niels Bohr Institute, Denmark) \\
Marika Asgari (University of Edinburgh, UK)\\
Mario Ballardini (Alma Mater Studiorum Universit\`a di Bologna, Italy)\\
Spyros Basilakos (Academy of Athens and Nat. Observatory of Athens, Greece) \\
Elia Battistelli (Sapienza Universit\`a di Roma and INFN sezione di Roma, Italy)\\
Micol Benetti (Universit\`a degli Studi di Napoli Federico II and INFN sezione di Napoli, Italy)\\
Simon Birrer (Stanford University, USA)\\
Fran\c{c}ois R. Bouchet (Institut d'Astrophysique de Paris, CNRS \& Sorbonne University, France) \\
Marco Bruni (Institute of Cosmology and Gravitation, Portsmouth, UK, and INFN Sezione di Trieste, Italy)\\
Erminia Calabrese (Cardiff University, UK)\\
David Camarena (Federal University of Espirito Santo, Brazil) \\
Salvatore Capozziello (Universit\`a degli Studi di Napoli Federico II, Napoli, Italy) \\
Angela Chen (University of Michigan, Ann Arbor, USA)\\
Jens Chluba (JBCA, University of Manchester, UK)\\
Anton Chudaykin (Institute for Nuclear Research, Russia) \\
Eoin \'O Colg\'ain (Asia Pacific Center for Theoretical Physics, Korea) \\
Francis-Yan Cyr-Racine (University of New Mexico, USA) \\
Paolo de Bernardis (Sapienza Universit\`a di Roma and INFN sezione di Roma, Italy) \\
Javier de Cruz P\'erez (Departament FQA and ICCUB, Universitat de Barcelona, Spain)\\
Jacques Delabrouille (CNRS/IN2P3, Laboratoire APC, France \& CEA/IRFU, France \& USTC, China)\\
Celia Escamilla-Rivera (ICN, Universidad Nacional Aut\'onoma de M\'exico, Mexico) \\
Agn\`es Fert\'e (JPL, Caltech, Pasadena, USA)\\
Fabio Finelli (INAF OAS Bologna and INFN Sezione di Bologna, Italy) \\
Wendy Freedman (University of Chicago, Chicago IL, USA)\\
Noemi Frusciante (Instituto de Astrof\'isica e Ci\^encias do Espa\c{c}o, Lisboa, Portugal)\\
Elena Giusarma (Michigan Technological University, USA) \\
Adri\`a G\'omez-Valent (University of Heidelberg, Germany)\\
Will Handley (University of Cambridge, UK) \\
Ian Harrison (JBCA, University of Manchester, UK) \\
Luke Hart (JBCA, University of Manchester, UK)\\
Alan Heavens (ICIC, Imperial College London, UK)\\
Hendrik Hildebrandt (Ruhr-University Bochum, Germany)\\
Daniel Holz (University of Chicago, Chicago IL, USA)\\
Dragan Huterer (University of Michigan, Ann Arbor, USA)\\
Mikhail M. Ivanov (New York University, USA) \\
Shahab Joudaki (University of Oxford, UK and University of Waterloo, Canada) \\
Marc Kamionkowski (Johns Hopkins University, Baltimore, MD, USA) \\
Tanvi Karwal (University of Pennsylvania, Philadelphia, USA) \\
Lloyd Knox (UC Davis, Davis CA, USA)\\
Suresh Kumar (BITS Pilani, Pilani Campus, India) \\
Luca Lamagna (Sapienza Universit\`a di Roma and INFN sezione di Roma, Italy) \\
Julien Lesgourgues (RWTH Aachen University) \\
Matteo Lucca (Universit\'e Libre de Bruxelles, Belgium)\\
Valerio Marra (Federal University of Espirito Santo, Brazil) \\
Silvia Masi (Sapienza Universit\`a di Roma and INFN sezione di Roma, Italy) \\
Sabino Matarrese (University of Padova and INFN Sezione di Padova, Italy) \\
Arindam Mazumdar (Centre for Theoretical Studies, IIT Kharagpur, India) \\
Alessandro Melchiorri (Sapienza Universit\`a di Roma and INFN sezione di Roma, Italy)\\
Olga Mena (IFIC, CSIC-UV, Spain)\\
Laura Mersini-Houghton (University of North Carolina at Chapel Hill, USA) \\
Vivian Miranda (University of Arizona, USA) \\
Cristian Moreno-Pulido (Departament FQA and ICCUB, Universitat de Barcelona, Spain)\\
David F. Mota (University of Oslo, Norway) \\
Jessica Muir (KIPAC, Stanford University, USA)\\
Ankan Mukherjee (Jamia Millia Islamia Central University, India) \\
Florian Niedermann (CP3-Origins, University of Southern Denmark) \\
Alessio Notari (ICCUB, Universitat de Barcelona, Spain) \\
Rafael C. Nunes (National Institute for Space Research, Brazil)\\
Francesco Pace (JBCA, University of Manchester, UK)\\
Andronikos Paliathanasis (DUT, South Africa and UACh, Chile) \\
Antonella Palmese (Fermi National Accelerator Laboratory, USA) \\
Supriya Pan (Presidency University, Kolkata, India)\\
Daniela Paoletti (INAF OAS Bologna and INFN Sezione di Bologna, Italy)\\
Valeria Pettorino (AIM, CEA, CNRS, Universit\'e Paris-Saclay, Universit\'e de Paris, France) \\
Francesco Piacentini (Sapienza Universit\`a di Roma and INFN sezione di Roma, Italy)\\
Vivian Poulin (LUPM, CNRS \& University of Montpellier, France) \\
Marco Raveri (University of Pennsylvania, Philadelphia, USA) \\
Adam G. Riess (Johns Hopkins University, Baltimore, USA) \\
Vincenzo Salzano (University of Szczecin, Poland)\\
Emmanuel N. Saridakis (National Observatory of Athens, Greece)\\
Anjan A. Sen (Jamia Millia Islamia Central University New Delhi, India) \\
Arman Shafieloo (Korea Astronomy and Space Science Institute (KASI), Korea)\\
Anowar J. Shajib (University of California, Los Angeles, USA) \\
Joseph Silk (IAP Sorbonne University \& CNRS, France, and Johns Hopkins University, USA)\\
Alessandra Silvestri (Leiden University, NL)\\
Martin S. Sloth (CP3-Origins, University of Southern Denmark) \\
Tristan L. Smith (Swarthmore College, Swarthmore, USA)\\ 
Joan Sol\`a Peracaula (Departament FQA and ICCUB, Universitat de Barcelona, Spain)\\
Carsten van de Bruck (University of Sheffield, UK) \\
Licia Verde (ICREA, Universidad de Barcelona, Spain)\\
Luca Visinelli (GRAPPA, University of Amsterdam, NL) \\
Benjamin D. Wandelt (IAP Sorbonne University \& CNRS, France, and CCA, USA) \\
Deng Wang (National Astronomical Observatories, CAS, China) \\
Jian-Min Wang (Key Laboratory for Particle Astrophysics, IHEP of the CAS, Beijing, China) \\
Anil K. Yadav (United College of Engg. \& Research, GN, India)\\
Weiqiang Yang (Liaoning Normal University, Dalian, China) \\

\noindent {\large \bf Abstract:} 
A precise measurement of the curvature of the Universe is of primeval importance for cosmology since it could not only confirm the paradigm of primordial inflation but also help in discriminating between different early Universe scenarios. The recent observations, while broadly consistent with a spatially flat standard $\Lambda$ Cold Dark Matter ($\Lambda$CDM) model, are showing tensions that still allow (and, in some cases, even suggest) a few percent deviations from a flat universe. In particular, the Planck Cosmic Microwave Background power spectra, assuming the nominal likelihood, prefer a closed universe at more than 99\% confidence level. While new physics could be in action, this anomaly may be the result of an unresolved systematic error or just a statistical fluctuation. However, since a positive curvature allows a larger age of the Universe, an accurate determination of the age of the oldest objects provides a smoking gun in confirming or falsifying the current flat $\Lambda$CDM model.

\begin{wrapfigure}{R}{0.45\textwidth}
%\begin{figure*}
\centering
\includegraphics[width=0.4\textwidth]{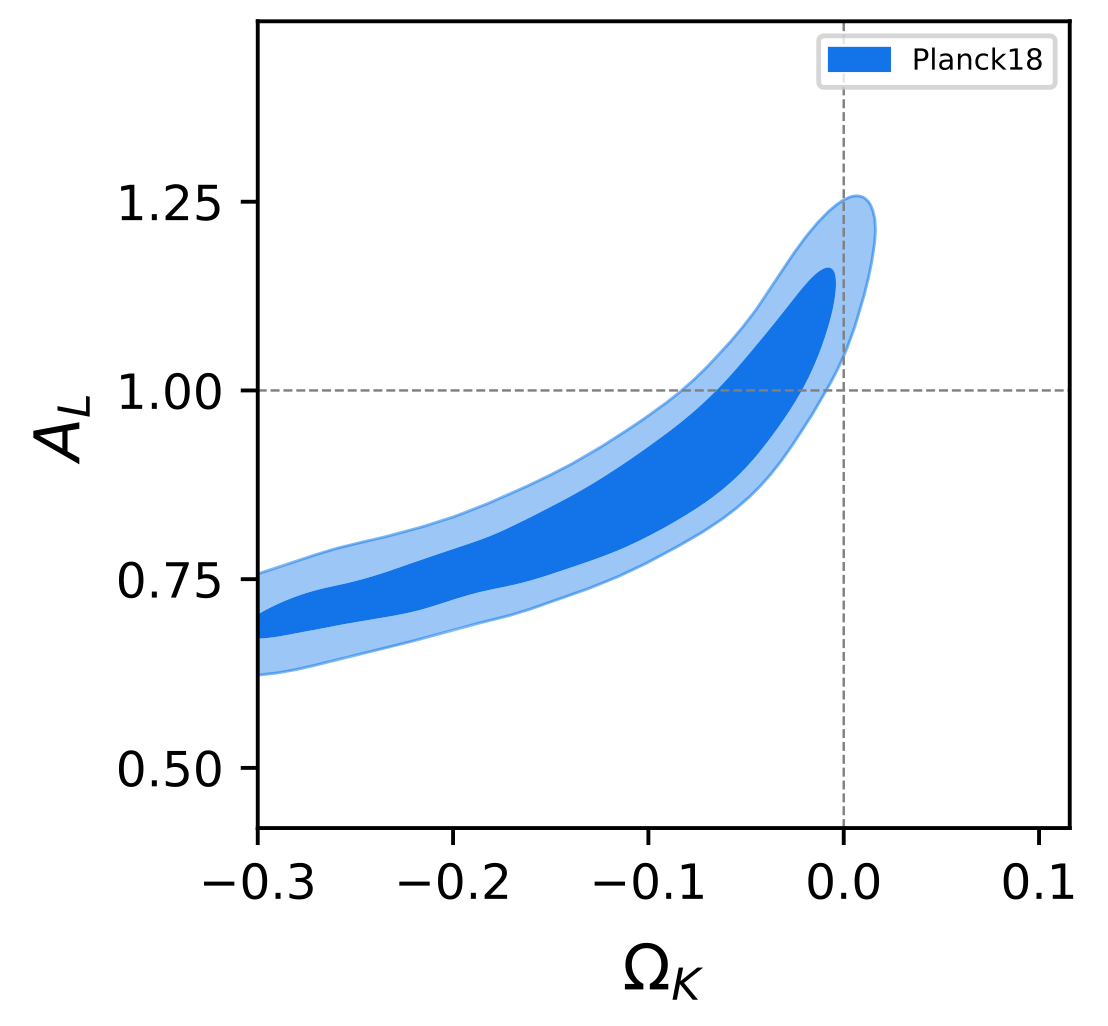}
\caption{68\% CL and 95\% CL contour plots for $\Omega_k$ and $A_{L}$ (from Ref.~\cite{DiValentino:2019qzk}).}
\label{2D}
%\end{figure*}
\end{wrapfigure}

\clearpage
\noindent {\bf The curvature of the Universe --} 
The flat $\Lambda$ Cold Dark Matter ($\Lambda$CDM) cosmological model describes incredibly well the current cosmological observations. However, together with the long standing {\it Hubble constant} $H_0$ disagreement~\cite{DiValentino:2020zio}, and $\sigma_8-S_8$ tension ~\cite{DiValentino:2020vvd}, there are some anomalies in the Planck 2018 cosmological results that deserve further investigations. Between them the most significant from the statistical point of view, is the preference at $3.4\sigma$ for a closed Universe~\cite{DiValentino:2019qzk,Aghanim:2018eyx,Handley:2019tkm}. Moreover, the Planck dataset also suggest an indication at more than $2\sigma$ for Modified Gravity~\cite{Aghanim:2018eyx,Ade:2015rim,Capozziello:2002rd}. This disagreement with the predictions for a flat universe of the standard model is connected with the higher, anomalous, lensing contribution in the Cosmic Microwave Background (CMB) power spectra, characterized by the $A_{L}$ parameter~\cite{Calabrese:2008rt,Aghanim:2018eyx}, that is strongly degenerate with $\Omega_k$ (see Fig.~\ref{2D}). A closed universe also solves a well-know tension above $2\sigma$ between the low and high multipoles regions of the angular power spectra ~\cite{Addison:2015wyg,Aghanim:2016sns,DiValentino:2019qzk}. This indication for curvature can be due to unresolved systematics in the Planck 2018 data, or can be simply due to a statistical fluctuation.

\begin{wrapfigure}{R}{0.45\textwidth}
%\begin{figure*}
\centering
\includegraphics[width=0.4\textwidth]{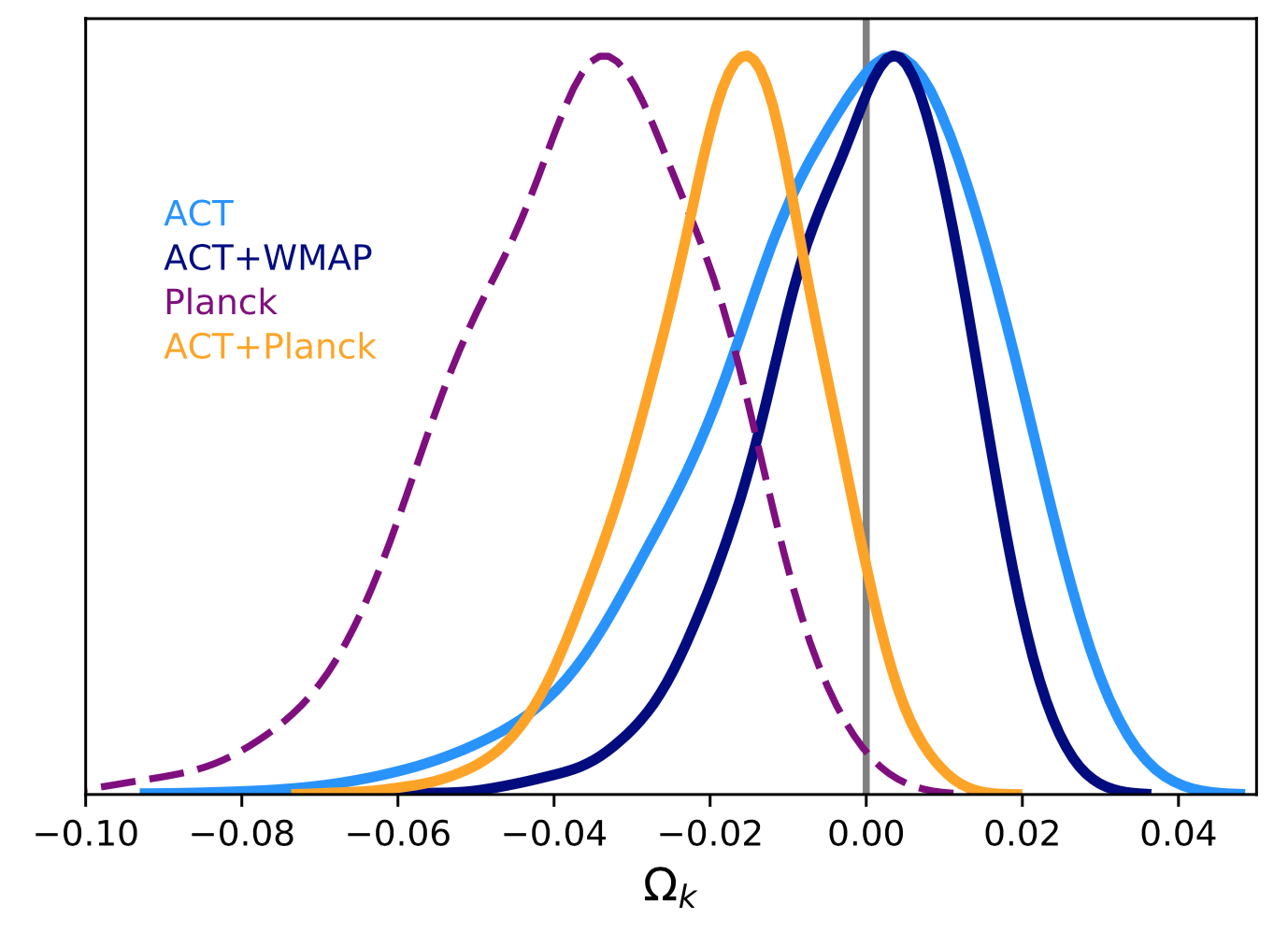}
\caption{1D posterior distributions on $\Omega_k$ (from Ref.~\cite{Aiola:2020azj}).}
\label{1D}
%\end{figure*}
\end{wrapfigure}

Indeed, while Planck 2018~\cite{Aghanim:2018eyx} finds $\Omega_k=-0.044^{+0.018}_{-0.015}$\footnote{All the bounds are reported at 68\% confidence level in the text.}, i.e. $\Omega_k<0$ at about $3.4\sigma$ ($\Delta \chi^2 \sim -11$), using the official baseline Plik likelihood~\cite{Aghanim:2019ame}, the evidence is reduced when considering the alternative CamSpec~\cite{Efstathiou:2019mdh} likelihood (see discussion in ~\cite{Efstathiou:2020wem}), albeit with the marginalized constraint still above the $99\%$ CL ($\Omega_k=-0.035^{+0.018}_{-0.013}$). Moreover, the recent results from the ground-based experiment ACT, in combination with data from the WMAP experiment, is fully compatible with a flat universe with $\Omega_k=-0.001^{+0.014}_{-0.010}$ 
while slightly preferring a closed universe when combined with a portion of the Planck dataset with $\Omega_k=-0.018^{+0.013}_{-0.010}$~\cite{Aiola:2020azj}(see Fig.~\ref{1D}). 
A closed universe is also preferred by a combination of non-CMB data made by Baryon Acoustic Oscillation (BAO) measurements~\cite{Beutler:2011hx,Ross:2014qpa,Alam:2016hwk}, supernovae (SNe) distances from the recent Pantheon catalog~\cite{Scolnic:2017caz}, and a prior on the baryon density derived from measurements of primordial deuterium~\cite{Cooke:2017cwo} assuming Big Bang Nucleosynthesis (BBN), but with a much larger $H_0$~\cite{DiValentino:2019qzk} completely in agreement with the SH0ES collaboration value R19~\cite{Riess:2019cxk}.
However, letting the curvature free to vary means to increase both the $H_0$ and the $S_8$ tensions~\cite{DiValentino:2019qzk}. 
Therefore, at the moment there are not theoretical models that can explain at the same time all the tensions and anomalies we see in the data. 
On the other hand, a flat universe is preferred also by Planck + BAO, or + CMB lensing~\cite{Aghanim:2018oex} or + Pantheon data. However these dataset combinations are in disagreement at more than $3\sigma$ when the curvature is free to vary~\cite{DiValentino:2019qzk,Handley:2019tkm}. In addition, though the error bars are so large that cannot discriminate between the models, a flat Universe is also in agreement with the analysis made by~\cite{Liu:2020pfa} using the $H(z)$ sample from the cosmic chronometers (CC) and the luminosity distance $D_L(z)$ from the 1598 quasars ($\Omega_k=0.08\pm0.31$) or the Pantheon sample ($\Omega_k=-0.02\pm0.14$), in agreement with the previous~\cite{Cai:2015pia}. Finally, in~\cite{Nunes:2020uex} a combination of BAO+BBN+H0LiCOW provides $\Omega_k=-0.07^{+0.14}_{-0.26}$ with $H_0$ in agreement with R19, while BAO+BBN+CC gives a positive $\Omega_k=0.28^{+0.17}_{-0.28}$.
In~\cite{Efstathiou:2020wem} it has been pointed out that is difficult to believe to a possible cosmological data conspiracy towards $\Omega_k=0$. However, a full agreement of the luminosity distance measurements, like Pantheon or R19, with Planck can be reached also ruling out both, a flat universe and a cosmological constant~\cite{DiValentino:2020hov}.

\noindent {\bf The Age of the Universe --} 
The age of the universe is an important piece of the puzzle because it connects $H_0$ and $\Omega_m$, both of which can be measured in the early and the late universe. The age is not just a prediction of the $\Lambda$CDM model, that for Planck 2018 is $t_U = 13.800 \pm0.024$ Gyr, but can also be measured using very old objects. For example, in~\cite{Valcin:2020vav} it is obtained $t_U = 13.35 \pm 0.16({\rm stat.})\pm0.5({\rm sys.})$ Gyr using populations of stars in globular clusters. Nevertheless, while robustness and accuracy tests have been done very extensively for CMB and quite extensively for BAO and SNe, for the age of the oldest objects we have to be more careful.
For example, one finds the ages of the oldest stars 2MASS J18082002–5104378 B equal to $t_* = 13.535 \pm 0.002$ Gyr~\cite{Schlaufman_2018}, but if the scatter among different models to fit for the age is taken into account the age becomes $t_* = 13.0 \pm 0.6$ Gyr~\cite{Jimenez:2019onw}, and the age of HD 140283 equal to $t_* = 14.46 \pm 0.8$ Gyr~\cite{Bond_2013}, but becomes $t_* = 13.5 \pm 0.7$ Gyr~\cite{Jimenez:2019onw} using the new Gaia parallaxes instead of original HST parallaxes.
Therefore, even if at present there is not real tension between the different $t_u$ determinations, most of the error-bars in the age determination comes from the fact that different stellar models do not really agree with each other at the required level of precision to be really able to help with the tensions in cosmology. Nevertheless, stellar models can/are expected to improve reducing this error significantly, and this could potentially unveil a tension on the age of the Universe. Trying to alleviate it by changing the Planck determination, would interestingly have an effect on the cosmological tensions.
A possibility to increase the age of the Universe, to be larger than the age of oldest stars, is by lowering the Hubble constant value, because of the anti-correlation between these parameters~\cite{deBernardis:2007jnf}.
For example, a positive curvature for the Universe, as suggested by Planck 2018, preferring a lower $H_0$ and worsening significantly the $H_0$ tension, predicts an older Universe $t_U = 15.31 \pm0.47$ Gyr.
Therefore, it seems that the only way to address the $H_0$ crisis if R19 is correct is to introduce an extremely recent (after $z\sim0.1$) departure from $\Lambda$CDM, requiring a great deal of fine-tuning.

\noindent {\bf Future --} 
Detecting a curvature $\Omega_k$ different from zero could be due to a local inhomogeneity biasing our bounds~\cite{Bull:2013fga}, and in this case CMB spectral distortions such as the KSZ effect and Compton-y distortions, present a viable method to constrain the curvature at a level potentially detectable by a next-generation experiment. If a curvature $\Omega_k$ different from zero is the evidence
for a truly superhorizon departure from flatness, this will have profound implication for a broad class of inflationary scenarios.
While open universe are easier to obtain in inflationary models~\cite{Gott82,Linde:1998gs,Bucher:1994gb,Kamionkowski:1994sv,Kamionkowski:1993cv}, with a fine-tuning at the level of about one percent one can obtain also a semi-realistic model of a closed inflationary universe~\cite{Linde:2003hc,PhysRevD.71.063502}. In~\cite{Leonard:2016evk} it has been shown that forthcoming surveys, even combined together, are likely to place constraints on the spatial curvature of $\sim 10^{-3}$ at 95\% CL at best, but enough for solving the current anomaly in the Planck data.
Experiments like Euclid and SKA, instead, may further produce tighter measurements of $\Omega_k$ by helping to break parameter degeneracies~\cite{DiDio:2016ykq,Vardanyan_2009}.

\noindent {\bf Summary --} 
In these four LoIs~\cite{DiValentino:2020vhf,DiValentino:2020zio,DiValentino:2020vvd} we presented a snapshot, at the beginning of the SNOWMASS process, of the concordance $\Lambda$CDM model and its connections with the experiment. This is a cutting-edge field in the area of cosmology, with unrestrained growth over the last decade. On the experimental side, we have learned that it is really important to have multiple precise and robust measurements of the same observable, with experiments conducted blind in regard to the expected outcome. This provides a unique opportunity to study similar physics from various points of view.
While on the theory side, it is really important having robust and testable predictions for the proposed physical models that can be probed with the data.
With the synergy between these two sides,
significant progress can be made to answer fundamental physics questions.
During the SNOWMASS process we plan to monitor the new advances on the field to come out with a clear roadmap for the coming decades.

\clearpage

%\noindent {\large \bf References:} (hyperlinks welcome) \\
\bibliographystyle{utphys}
\bibliography{H0}

\end{document}